# Protein Folding: From Classical Issues to a New Perspective

Jorge A. Vila

IMASL-CONICET, Universidad Nacional de San Luis, Ejército de Los Andes 950, 5700 San Luis, Argentina.

## Abstract

The Levinthal paradox exposes many critical questions on the protein folding problem, among which we could point out *why* proteins can reach their native state in a biologically reasonable time. A proper answer to this question is of foremost importance for evolutive biology since it enables us to understand life as we know it. Preliminary results, based on the upper bound protein marginal-stability limit, together with transition state theory arguments, lead us to show that two-state proteins must reach their native state in, at most, seconds rather than ($\sim 10^{27}$) years - as indicated by a naïve solution of the Levinthal paradox. This outcome -added to the amide hydrogen-exchange protection factors analysis- makes it possible for us to suggest *how* a protein point mutations and/or post-translational modifications impact its folding time scales but not its upper bound limit that obeys the physics ruling the process. Noteworthy for almost 50 years, the protein folding problem -as the Levinthal paradox- has been a topic of passionate debate because Anfinsen's challenge -*how* a sequence encodes its folding- remains unsolved despite the smashing success on accurately predicting the protein tridimensional structures by state-of-the-art numerical-methods. Aimed to unlock this long-standing challenge, we propose a new perspective of protein folding, specifically, as a problem that should be devised as an "analytic whole" -a Leibniz & Kant's notion-. This viewpoint might help us decode Anfinsen's challenge and, thus, open new avenues for future research in the protein folding field.

## Overlook

Evolution and protein folding are intertwined processes. Indeed, protein sequences, encoded by DNA, determine their tridimensional structure -Anfinsen (1973)- which in turn determines their function, while evolution could alter either one by mutations. Then, does the protein folding time restrain the mutation frequency? If this were the case, which is its impact on evolution? Whatever the answer to these questions, protein folding cannot happen in cosmic times ($\sim 10^{27}$ years), as foreseen by an exhaustive sampling of all possible conformations for a 100-residue protein (Zwanzig *et al*., 1992). In fact, proteins fold from milliseconds to seconds (Sali *et al*., 1994; Karplus, 1997). As the reader must be aware, several possible solutions to this apparent contradiction, also known as Levinthal's paradox (Levinthal, 1968), exist in the literature (Zwanzig *et al*., 1992; Dill & Chan, 1997; Karplus, 1997; Rooman *et al*., 2002; Ben-Naim, 2012; Finkelstein & Garbuzynskiy, 2013; Martinez, 2014; Ivankov & Finkelstein 2020). Therefore, we will not revisit this problem here. However, the existence of solutions to this paradox does not assure a clear answer to the following key question: *Why* can proteins reach their native state in a biologically reasonable time? As a strategy to answer this question, we will prove that a reasonable estimation of the height of the activation barrier (see Fig.1), separating the native state from the highest-energy nativelike conformation -beyond which the protein unfolds or becomes nonfunctional (Vila, 2019; 2022)- will enable us to determine the slowest folding time for two-state monomeric proteins. In this way, not only could we find an answer to the query, but we could also propose another solution to Levinthal's paradox. Before resuming the analysis, let us go back to the last question. Should we focus on *why* rather than *how* proteins reach their native state in a biologically reasonable time? This dilemma does not have a simple solution because both are relevant queries. Indeed, the interrogative *how* is associated with determining the mechanism, *e.g*., the routes or pathway/s of the unfolding/folding (Sali, *et al.,* 1994; Wolynes *et al*., 1995; Lazaridis & Karplus, 1997; Jackson, 1998; Lindorff-Larsen *et al*., 2011; Englander *et al*., 2014; Wolynes, 2015; Li & Gong, 2022), while the *why* is associated with identifying the main factors -independently of the mechanism- governing the unfolding (and folding) process. An attempt to answer *how* proteins reach their native state in a biologically reasonable time has been recently analyzed (Ivankov & Finkelstein 2020). Therefore, we choose to focus on *why* -rather than on *how*- two-state proteins reach their native state in a biologically reasonable time, among other reasons, because this analysis will enable us, firstly, to identify the physical nature of this protein feature and, secondly, to understand the origin of the folding time scale changes upon point mutations and/or post-translational modifications (Martin & Vila, 2020), regardless of the fold-class, chain length or amino-acid sequence.

Overall, from here on, we will focus on determining the nature of the main factors controlling the slowest time of the unfolding (and folding) process for a two-state monomeric protein, *i.e.*, an answer to *why* proteins reach their native state in a biologically reasonable time. Afterwards, we will clear up how such a slowest time caps the frequency of point mutations and, thus, how it could influence, among other factors, protein evolvability speed. Later, the nature of the protein unfolding (and folding) time scale changes upon a point mutation will be examined in terms of (*i*) the protein-marginal stability (Vila, 2019; Martin & Vila, 2021; Vila, 2022); (*ii*) arguments from the transition state theory (Ivankov & Finkelstein, 2020); and (*iii*) the amide hydrogen-exchange protection factors, a highly sensitive probe of the stability, structure, and folding of proteins (Hvidt & Linderstrøm-Lang, 1954; Berger *et al*., 1957; Privalov & Tsalkova, 1979; Englander *et al*., 1997; Huyghues-Despointes *et al*., 1999; Craig *et al*., 2011; Balasubramaniam & Komives, 2013; Persson & Halle, 2015). Finally, we will analyze why the protein folding problem remains unsolved -while their tridimensional structures can be forecast with high-precision (Jumper et al., 2021)- in terms of a notion from Leibniz & Kant (Gómez, 1998).

**I.- Two-State Protein Folding Time Scales**

Among the possible solutions to the time scales for protein folding, we distinguish three papers determining a plausible relation between protein length (*N*), with *N* being the number of residues, and the folding time logarithm (*ln* τ), namely, *ln* τ ~ $N^{1/2}$ (Thirumalai, 1995), ~ *ln* (*N*) (Gutin *et al*., 1996), and ~ $N^{2/3}$ (Finkelstein & Badretdinov, 1997; Wolynes, 1997). Although an analysis of such a relationship is vital, given the strongly observed anticorrelation -between *N* and *ln* τ- for three-state folding proteins (*R* ~ −0.80) (Galzitskaya *et al*., 2003), is it also equally important to highlight that such a relationship for two-state folding proteins is nearly inexistent (*R* ~ −0.07) (Galzitskaya *et al*., 2003). Therefore, we will focus on determining a plausible explanation for the latter. For this purpose, we will resolve the slowest unfolding (and folding) time ($\tau_{max}$) for a monomeric two-state protein in terms of the result obtained for the proteins' marginal-stability upper bound obtained via a statistical-mechanics analysis of the Partition Function in the thermodynamic limit, also known as the infinite chain limit (Vila, 2019; Vila, 2021). As a consequence of this analysis, the result obtained for $\tau_{max}$ will be valid not only for proteins of any length (*N*), but also for any amino-acid sequence or fold-class, in agreement with experimental pieces of evidence (Galzitskaya *et al*., 2003). Derivation of the expected value for $\tau_{max}$ will be valid under the following set of conjectures and facts:

1.- The word mutation could refer to either an amino-acid substitution in the protein sequence -as a result of a nucleotide pair replacement- or post-translational modifications -that could expand the diversity of the proteome by several orders of magnitude (Garay *et al*., 2016)-. Considering that each post-translational modification could be thought of as a different amino acid to the 20 naturally occurring, then a mutation will merely refer to a protein sequence modification (Martin & Vila, 2020).

2.- The two-state protein unfolding (and folding) model alludes to a process in which the thermodynamics and kinetic stability happen only between the native-state and unfolded states, which are separated by an energetic barrier higher than thermal fluctuation energy (Akmal & Muñoz, 2004; Kuwajima, 2020). In other words, folded and unfolded states are separated by an ensemble of high-energy set of structures, *i.e*., the Transition States Ensemble (TSE), representing the energetic barrier for the process (see Fig. 1) (Privalov, 1979; Matouschek *et al*., 1989; Itzhaki *et al*., 1995; Englander, 2000; Fersht & Daggett, 2002; Akmal & Muñoz, 2004; Shakhnovich, 2006). In this simple folding model, there are no stable intermediate states necessary to complete the process.

3.- The largest size of the Gibbs free-energy barrier ($\Delta$G) between the native-state and the highest point of the free-energy profile (see Fig. 1) is assumed to be given by the protein marginal-stability upper bound limit, *i.e*., $\Delta$G < ~ 7.4 kcal/mol, which (*i*) is a universal feature of proteins, *i.e*., was obtained regardless of the fold-class or its amino-acid sequence (Vila, 2019); (*ii*) is a consequence of Anfinsen's dogma validity (Vila, 2019; 2021); and (*iii*) represents a threshold beyond which a conformation will unfold and become nonfunctional (Martin & Vila, 2020; Vila, 2021; 2022). This assumption offers a rationale for the conjecture that the height of the thermodynamic energetic barrier -and, hence, its existence- is due to the validity of the thermodynamic hypothesis or Anfinsen dogma (Vila, 2019; 2021; 2022). In other words, the physics that rule the folding.

4.- The unfolding (and folding) approach for monomeric two-state proteins is a reversible thermodynamic driven process (Privalov, 1979; Matouschek *et al*., 1989).

5.- It is assumed that point mutations *mainly affect* the native-state stability (Zeldovich *et al*., 2007). This assumption is, firstly, equivalent to assuming an average $\phi$-value -a technique commonly used to examine the kinetic effects on the protein folding upon a point mutation (Matouschek *et al*., 1989; Itzhaki *et al*., 1995; Campos, 2022)- closer to ~0 than ~1; and, secondly, in line with the results of the $\phi$-value analysis for more than 800 mutations for 24 two-state proteins showing a $<\phi> \sim 0.24$ (Naganathan & Muñoz, 2010).

6.- The unfolding Gibbs free energy ($\Delta G_U$) between the wild-type (*wt*) and the mutant (*m*) protein can be effortlessly computed as $\Delta\Delta G_U = (\Delta G_U^m - \Delta G_U^{wt})$ (Bigman & Levy, 2018). This definition -together with assumption 5- enables us to propose (Vila, 2022) a reasonable strategy to assess the change in the protein marginal-stability upon point mutations ($\Delta\Delta G$), namely as $\Delta\Delta G \sim \Delta\Delta G_U$.

7.- The speed limits ($\tau_0$) of two-state protein' unfolding (barrier-less limit) have been discussed in great depth in the literature (Zana, 1975; McCammon, 1996; Hagen *et al*., 1996; Mayor *et al*., 2000; Krieger *et al*., 2003; Yang & Gruebele, 2003; Akmal *et al*., 2004; Muñoz *et al*., 2008; Ivankov & Finkelstein, 2020; Glyakina & Galzitskaya, 2020; Muñoz & Cerminara, 2016; Chung & Eaton, 2018; Eaton, 2021), and there is a consensus that it should be within the following range of values:

$$\sim 10^{-8}\,[\text{sec}] < \tau_0 < \sim 10^{-5}\,[\text{sec}] \tag{1}$$

Let us briefly illustrate the impact of these folding times limits on protein evolvability. If a given 100-residue two-state protein cannot fold faster than $\tau_0 \sim 10^{-8}$ (or $\sim 10^{-5}$) seconds, and if life began on earth around a billion ($\sim 10^9$) years ago, its Protein Space size (Maynard Smith, 1970) would contain at most $\sim 10^{24}$ (or $\sim 10^{21}$) sequences. If this were the case, the average-mutation rate per amino acid ($\xi$) should be $\leq \sim 1.74$ (or $\leq \sim 1.62$) since $\xi$ must satisfy $\xi^{100} = \sim 10^{24}$ (or $\sim 10^{21}$). The fact that $\xi < 2$ is of paramount importance, from an evolutive point of view, because it means that only a fraction of a given protein sequence is available for a mutation at any one time, in agreement with both previous estimations of the Protein Space (PS) size (Vila, 2020) and existent pieces of evidence (Margoliash & Smith, 1965; Sarkisyan *et al*, 2016). A detailed discussion of an accurate estimation of the PS size in light of the factors that govern it is of crucial importance from an evolutive point of view (Mandecki, 1998; Dryden *et al*. 2008; Romero & Arnold, 2009; Ivankov, 2017). Moreover, its solution shows that evolution and folding time scales are interwoven phenomena since a reliable estimation of the PS size in light of molecular evolution is not just a combinatorial problem (Vila, 2020).

8.- The time ($\tau$) to overcome the free-energy barrier $\Delta G$ (shown in Fig.1) may be computed by using an argument from the transition state theory, as (Ivankov & Finkelstein, 2020):

$$\tau = \tau_0 \exp{(\beta \, \Delta G)} \; [sec] \tag{2}$$

where the lower and upper bound of the pre-exponential factor ($\tau_0$) is given in Eq. (1), $\beta = 1/RT$, $R$ is the gas constant and $T$ is the absolute temperature (298K for all the calculations).

9.- We will focus our attention on the unfolding rather than on the folding process because the former enables us to make a quickly estimation of the height of the Gibbs free-energy difference ($\Delta G$) between the native-state (representing a well-defined reference-point) and the highest point of the Transition States Ensemble (TSE) (see Fig.1), a point beyond which a protein becomes unfolded or nonfunctional (Vila, 2019; 2022). Due to the principle of microscopic reversibility, the TSE should be the same for the unfolding and folding processes, *e.g*., as shown by the analysis of the rates and equilibria of folding from ~100 mutants strategically distributed throughout the protein chymotrypsin inhibitor 2 (Itzhaki *et al*., 1995). This conjecture is in line with the observed folding/unfolding data from 108 proteins (70 showings two-state kinetics) that demonstrate that the logarithm of the folding and unfolding times is well correlated ($R$ ~0.8) and that such correlation is better for a two than that multiple-state proteins (Glyakina & Galzitskaya, 2020). If the free energy barrier vanishes ($\Delta G$ ~ 0), a downhill, barrierless or one-state unfolding (Garcia-Mira *et al.*, 2002; Naganathan *et al.*, 2005; Muñoz *et al*., 2008) occurs in times given by ($\tau_0$).

10.- After assuming the validity of all the above conjectures and facts, it is possible to determine from Eq. (2) (with $\Delta G \sim 7.4$ kcal/mol and $\tau_0$ given by Eq. 1) the following range of $\tau_{max}$ values:

$$\sim 10^{-3} \text{ [sec]} \leq \tau_{max} \leq \sim 1 \text{ [sec]} \tag{3}$$

This range of values for $\tau_{max}$ is certainly in agreement with observed protein folding times under biological conditions (Garbuzynskiy *et al*., 2013). Indeed, results from 65 two-state proteins that fold in aqueous solution without folding intermediates show a folding time upper bound limit $\leq \sim 10$ [sec] (Garbuzynskiy *et al*., 2013; Ivankov & Finkelstein, 2020). The range of variation for $\tau_{max}$ shown in Eq. (3) depends neither on the amino-acid sequence nor on chain length for a two-state protein, in line with the observation that chain length has an almost null correlation ($R \sim -0.07$) with the folding time logarithm (Plaxco *et al*., 2000; Galzitskaya *et al*., 2003). Moreover, the result for $\tau_{max}$ given by Eq. (3) solves a critical inquiry of Levinthal's paradox: how long it takes for a protein to reach its native state.

A detailed review of the use of computational methods aimed at predicting protein folding times falls outside of the scope of this article. For those interested in the analysis of different machine learning algorithms used for this purpose as the pros and cons of each of them, a recent review on this issue is highly recommended (Ivankov & Finkelstein, 2020).

## II.- Protein Folding Time Scale Changes upon Point Mutations

If the free-energy barrier height ($\Delta G$) rules the unfolding (and folding) time $\tau$ for a two-state protein, then a single point mutation could affect it by either increasing (stabilizing) or decreasing (destabilizing) the marginal-stability (Vila, 2022), but this should always happen within the allowed range for the $\tau$ values, namely, $\tau_0 < \tau < \tau_{max}$. Indeed, there are numerous pieces of evidence supporting this. Let us start by examining the physics that rules the phenomenon of folding time changes upon a protein point mutation. The ratio between the wild-type protein folding time ($\tau_{wt,f}$) and that of this protein upon a point mutation ($\tau_{m,f}$) can be computed using Eq. (2), as (Chaudhary *et al.*, 2015; Ivankov & Finkelstein, 2020):

$$\Delta\tau_m = (\tau_m/\tau_{wt}) = \exp(\beta\,\Delta\Delta G_m) \quad \Rightarrow \quad RT\,ln\,\Delta\tau_m = \Delta\Delta G_m \qquad (4)$$

where $\Delta\Delta G_m = (\Delta G_m - \Delta G_{wt}) \sim \Delta\Delta G_U$ is the change, upon a single-point mutation, between the mutant and the wild-type Gibbs free-energy gap ($\Delta G$), respectively (Vila, 2022). Therefore, the following relationship for a two-state protein should also hold (Vila, 2022) $\Delta\Delta G_m = RT$ ln ($\Delta P_m$), where $\Delta P_m = (P_{f,m} / P_{f,wt})$ and $P_{f,m}$ and $P_{f,wt}$ are the corresponding protection factors ($P_f$) for the mutant (*m*) and the wild-type (*wt*) protein, respectively (Vila, 2021, 2022). Note that $P_f$ represents the resistance of the amide Hydrogen-eXchange (HX) in the native state relative to that of the highest free-energy conformation in the ensemble of folded states (Vila, 2021, 2022), *i.e.*, in the Transition States Ensemble -TSE- shown in Figure 1. This is a reasonable conjecture because our interest focuses on a particular region of the conformational space, namely, the one in which the EX2 limit is valid (Bahar *et al.*, 1998). Hence,

$$\Delta\Delta G_m = RT\,ln\,(\Delta\tau_m) = RT\,ln\,(\Delta P_m) \qquad (5)$$

This equation describes the change in the unfolding (and folding) time logarithm upon a point mutation (*ln* $\Delta\tau_m$) and how it corresponds with variations happening in the ensemble of conformations coexistent with the native state (*ln* $\Delta P_m$). This result is in line with evidence indicating a good correlation (*R* ~ 0.8) -observed for 108 proteins- between the logarithm of the unfolding rates and protein stabilities (Glyakina & Galzitskaya, 2020). Let us illustrate its significance with an example. A database of 790 mutants for 26 two-state proteins shows that the changes of *ln* $\Delta\tau$ are in the range ~ −5.2 to ~2.6 (Chaudhary *et al.*, 2015). Consequently, the $\Delta\Delta G$ changes should be within the range ~ −3.1 to ~1.5 kcal/mol (see Figure 2). Since most of these free-energy changes (77%) are destabilizing ($\Delta\Delta G < 0$), a well-known effect of protein mutations (Zeldovich *et al.*, 2007; Tokuriki & Tawfik, 2009; Arnold, 2009; Socha & Tokuriki, 2013), let us analyze their impact on folding time changes in terms of the TSE structural fluctuations. If $\Delta\Delta G$ ~ −3.1 kcal/mol, then $P_{f,m} \sim P_{f,wt}$ x10$^{-2}$. This result implies that the native state of the mutant (*m*) should be significantly less resistant to the amide HX than that of the wild-type (*wt*). Therefore, a significant reduction of the height of the energetic barrier and, hence, the range of structures populating the TSE together with a rearrangement of their relative free energies should happen (Lazaridis & Karplus, 1997; Vendruscolo *et al*., 2003). Consequently, the protein unfolding upon mutation would be (~$10^2$ times) faster than that of the wild-type, as inferred from equation (5).

It is worth noting that equation (5) will be valid after $j$ ($j > 1$) consecutive point mutation steps -because ΔG is a state function- and, hence, it can be generalized straightforwardly by replacing $m \rightarrow j$ (Vila, 2022). This generalization is relevant if, for example, we model evolution as a walk in the Protein Space, *i.e*., one where "…*if evolution by natural selection is to occur, functional proteins must form a continuous network which can be traversed by unit mutational steps without passing through nonfunctional intermediates*…" (Maynard Smith, 1970).

Overall, the above analysis confirms that evolution influences -through mutations- the unfolding (and folding) time scales, *i.e*., by altering the height and composition of the energetic barrier, but not their rate limiting step set by the physics of folding, namely, by the largest-possible change in the free energy barrier (ΔΔG < ~ 7.4 kcal/mol). This barrier defines from a thermodynamic standpoint a threshold beyond which a two-state protein unfolds or becomes non-functional and, from a kinetic viewpoint, the time ceiling for the unfolding process.

## III.- Why does the Protein Folding Problem Remain Unsolved?

The protein folding problem has been under investigation since Anfinsen announced his thermodynamic hypothesis almost 50 years ago (Anfinsen, 1973). Despite the enduring efforts to explain this biological process (Lewis *et al*., 1971; Kuntz, 1972; Tanaka & Scheraga, 1975; Anfinsen & Scheraga, 1975; Némethy & Scheraga, 1977; Creighton, 1978; Rossmann & Argos, 1981; Go, 1983; Dill, 1990; Dill & Chan, 1997; Dobson, 2003; Onuchic & Wolynes, 2004; Rose *et al*., 2006; Shakhnovich, 2006; Dill *et al*., 2007; Dill *et al*., 2008; Dill & MacCallum, 2012; Wolynes, 2015; Nassar *et al*., 2021; Finkelstein, 2018; Rose, 2021), the Anfinsen challenge: *how* a sequence encodes its folding -essential to understanding both protein function or malfunction-remains unknown (Cramer, 2021; Clementi, 2021; Jones & Thornton, 2022), even when its tridimensional structure can be predicted with high-accuracy by state-of-the-art numerical-methods (Jumper *et al*., 2021). Regrettably, Anfinsen's challenge might continue to be unsolved in the foreseeable future, and the reason is as follows. The field concept proved most successful and led to the formulations of several outstanding physical problems, from Maxwell's equations to the theory of relativity (Einstein & Infeld, 1961). Following such a concept, the protein folding problem has been attempted to solve since ~1965 (Némethy & Scheraga, 1965; Gibson & Scheraga, 1967; Scheraga, 1968; Lifson & Warshel, 1968; Momany et al., 1975) by using various force-fields, which share the feature that they possess very similar functional forms (Best, 2019; Arnautova *et al*., 2005). However, this approach has not allowed us, for almost 60 years, to accurately determine -at atomic-level and from random conformations- the tridimensional structure of proteins, except for a few cases, *e.g*., for protein A (Vila *et al*., 2003). The cause for this long-lasting failure to crack Anfinsen's challenge could be explained with the help of the following idea: "the whole is more than the sum of the parts." This comes from Leibniz & Kant's notion of space (and time), devised as "analytic wholes", *i.e*., the one where "…*its priority makes it impossible to obtain it by the additive synthesis of previously existing entities*…" (Gómez, 1998). From this viewpoint the protein folding problem should be understood as an "analytic whole". In terms of this analogy, forecasting the tridimensional protein structure using either force-fields or state-of-the-art machine learning methods, such as AlphaFold2 (Jumper *et al*., 2021), can be understood as an attempt to reach the whole, naturally with different degrees of success (Kryshtafovych *et al*. 2021; Marx, 2022). However, despite the smashing success in accurately predicting the tridimensional protein structures of methods such as AlphaFold2, several unsolved issues, *e.g.*, protein misfolding organization (Serpell *et al*., 2021), protein pathway (Jones & Thornton, 2022), an accurate determination of structural and marginal-stability changes upon protein point-mutations and/or post-translational modifications (Pancotti *et al*., 2022; Serpell *et al*., 2021; Buel *et al*., 2022; Vila, 2022), predicting specific, nonspecific, or high-order epistasis effects (Domingo *et al*., 2019; Miton *et al*., 2020), *etc*., are, conceivably, pieces of evidence suggesting that the protein folding problem should indeed be studied as an "analytic whole", but not as the "whole" understood as a sum of parts. Therefore, its solution demands treating protein folding as an $N$-body problem (being $N$ the number of amino-acids of the sequence) and, consequently, through a function that cannot break down in terms of lower order functions. As an alternative to surmount this colossal problem are strategies considering a sum of $k$-body functions ($2 \leq k \leq N$), in which each cannot be expressed as a sum of lower order functions either (Wang *et al*., 2021). All of the above should not be surprising since existing evidence indicates that "…*many-body interactions can play a significant role in governing the folding mechanisms of two-state proteins when described at the residue level*…" (Ejtehadi *et al*., 2004). We should recall that a protein native state is marginally stable (Privalov &Tsalkova, 1979; Albert, 1989; Vila, 2019) and, hence, stabilized by weak interactions, such as those due to the interplay of pairwise and many-body interactions on both the proteins and the solvent (Martin & Vila, 2020). At this point, it is worth mentioning the expression attributed to Gustave Flaubert: "*Le bon Dieu est dans le detail.*"

Overall, the protein folding problem should not be approached by methods entirely based on pairwise additive interactions because it should be devised as an "analytic whole" as Leibniz and Kant's notion of space (and time). Overcoming this formidable challenge might explain the immense number of parameters ($\sim 10^7$) needed to characterize residues or blocks of residues in methods such as AlphaFold (Torrisi *et al.*, 2020). A description of the main differences between machine-learning-based and force-field-based approaches to solving the protein folding problem or how AlphaFold works is beyond the scope of this manuscript. Yet, an attempt to clarify these issues can be found in recently published articles (Skolnick *et al.*, 2021; Fersht, 2021; Jones & Thornton, 2022; Baek & Baker, 2022).

## Conclusions

Firstly, the analysis presented here has enabled us to show a simple solution to Levinthal's paradox, which originates in a crucial question: how long does it take for a protein to reach its native state? As proved, it takes seconds, not years, for a two-state protein to reach its native state. Moreover, we have been able to determine the physics behind the folding time scale changes upon point mutations and/or post-translational modifications and *how* the folding rate restricts, among many other factors, the protein evolvability. In addition, the analysis enabled us to understand *why* proteins reach their native state in a biologically reasonable time. Specifically, because the largest-possible change in the two-state protein free-energy barrier ($\sim 7.4$ kcal/mol) is a consequence of the thermodynamic hypothesis validity -or Anfinsen's dogma- *i.e.*, a limit set by the physics of folding, regardless of the fold-class, chain length or amino-acid sequence.

Finally, despite the fantastic progress in the structural and evolutive biology fields, an accurate answer to *how* a sequence encodes its folding remains unsolved, perhaps, because the folding problem should be devised as an "analytic whole". This new perspective, far from being gloomy, could help us decide where to direct our effort to find a final solution to Anfinsen's long-lasting challenge and be aware of the origin and limitations of several existing approaches to solve it based primarily on pairwise additive interactions.

## Acknowledgments

The author acknowledges support from the IMASL-CONICET-UNSL and ANPCyT (PICT-02212), Argentina.

## Dedication

Dedicated in memoriam to Harold A. Scheraga, a world-class scientist, and an outstanding mentor, colleague, and friend (Vila, 2020b).

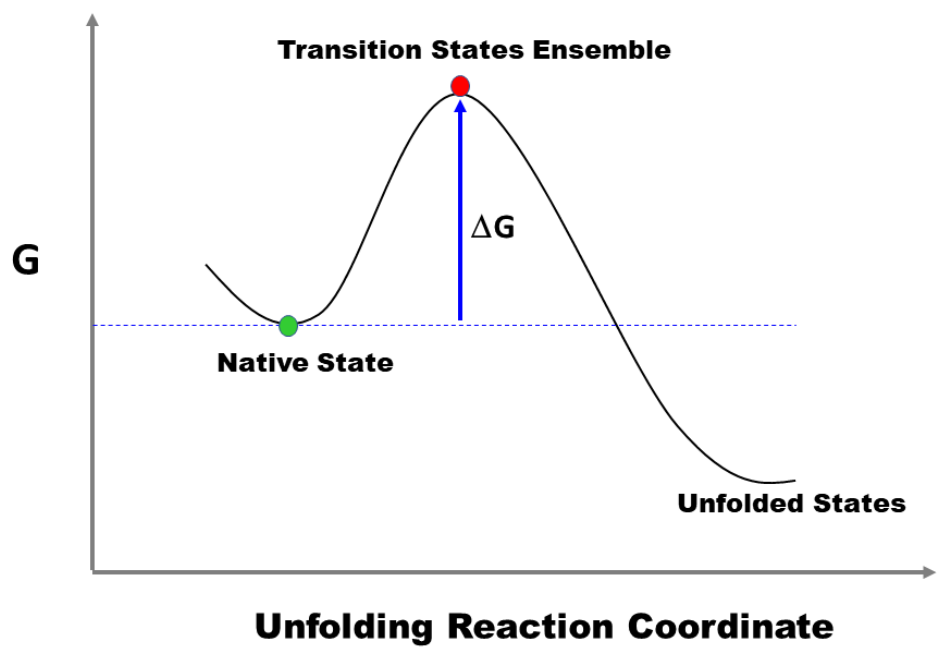

**Figure 1**. Cartoon of the Gibbs free-energy profile (**G**) for a two-state protein unfolding. The native-state and the highest point of the free-energy profile are highlighted as green- and red-filled dots, respectively. The Gibbs free-energy gap between these two states is indicated by **ΔG**.

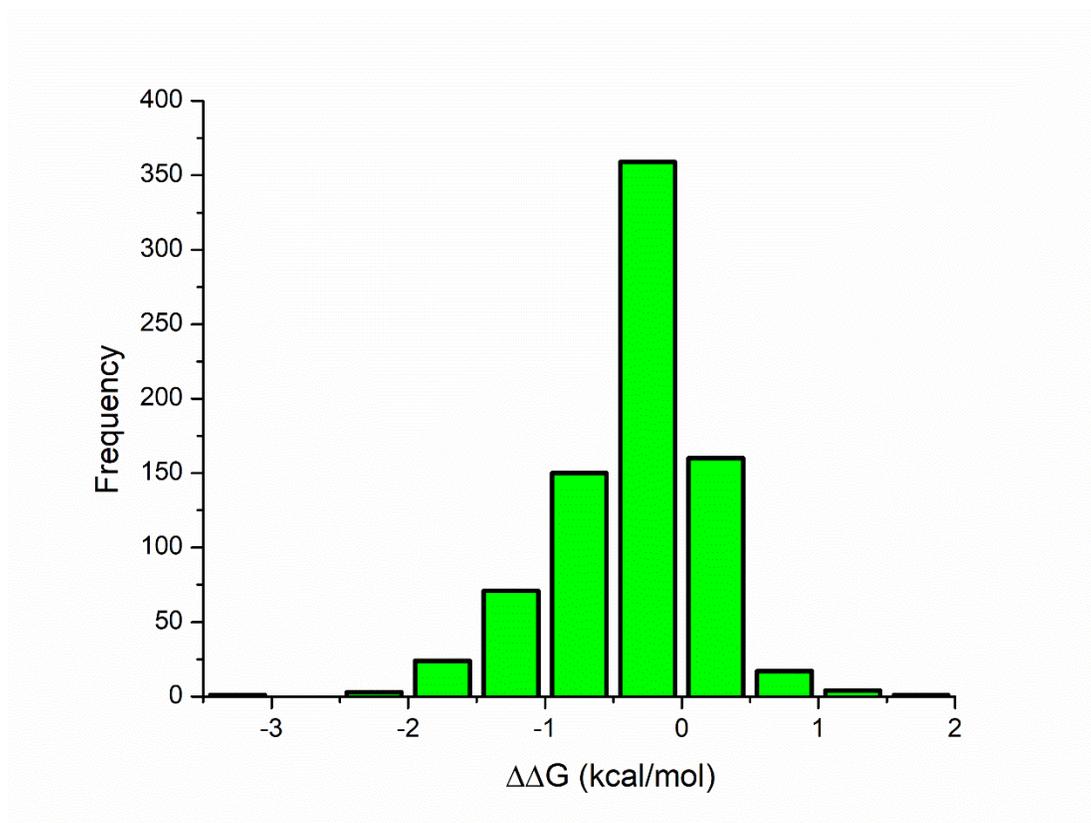

**Figure 2.-** A plot of the free-energy barrier changes ($\Delta\Delta G$), computed from a data set containing the observed folding time logarithm for 790 mutants from 26 two-state proteins. (Chaudhary *et al*, 2015). The distribution shows that most of those mutations, namely 608 out of 790, are destabilizing, *i.e.* $\Delta\Delta G < 0$.